\documentstyle[preprint,prl,aps,epsf]{revtex}

\begin{document}
\title{Nuclear Shadowing and High-$p_T$ Hadron Spectra
in Relativistic Heavy Ion Collisions}

\author{S. R. Klein$^1$ and R. Vogt$^{1,2}$}
 
\address{{
$^1$Nuclear Science Division, Lawrence Berkeley National Laboratory, 
Berkeley, CA 94720, USA\break 
$^2$Physics Department, University of California, Davis, CA 95616, USA}\break}
 
\vskip .25 in
\maketitle
\begin{abstract}

We explore how nuclear modifications to the nucleon parton
distributions affect production of high-$p_T$ hadrons in heavy ion
collisions.  We perform a leading order calculation of the high-$p_T$
charged pion, kaon and proton spectra using standard fragmentation functions
and shadowing parameterizations.  We also consider alternate models of
shadowing.  Near midrapidity, shadowing is a small effect and
cannot explain the large observed suppression of high-$p_T$ hadrons.
We also consider the isospin difference between protons and nuclei and 
find that it is also a small effect on high-$p_T$
hadron production.

\end{abstract}
\narrowtext

Some of the most interesting experimental results to come out of the
Relativistic Heavy Ion Collider (RHIC) at Brookhaven National
Laboratory have involved the observed suppression of high-$p_T$ hadron
production in collisions of gold nuclei at a center of mass energy,
$\sqrt{S_{NN}}$,  of 130 and 200 GeV per nucleon.  Both the PHENIX\cite{phenix}
and STAR\cite{star} collaborations have published results showing that
production of hadrons with transverse momentum $p_T>2$ GeV/$c$ are
suppressed compared to the $pp$ reference spectrum convoluted with the
number of binary collisions.  The suppression is given in terms of the
ratio
\begin{equation}
R_{AA}(p_T)  = {d \sigma_{AA}/dp_T \over T_{AA} d\sigma_{pp}/dp_T }
\end{equation}
where $d\sigma_{AA}/dp_T$ and $d\sigma_{pp}/dp_T$ are the hadron $p_T$
distributions in $AA$ and $pp$ collisions respectively and $T_{AA}$ is
the nuclear overlap function.  The product of the inelastic
nucleon-nucleon cross section with the nuclear overlap function is the
number of binary nucleon-nucleon collisions in a given impact
parameter range.

The data show that, for $p_T>2$ GeV/$c$, $R_{AA}(p_T)$ is much less
than 1.  For charged hadrons (pions, protons and kaons),
$R_{AA}(p_T)\approx 0.4$\cite{phenix,star}, while for $\pi^0$,
$R_{AA}(p_T)\approx 0.3$\cite{phenix}.  It appears that, for protons,
there is very little suppression, with $R_{AA}(p_T)\approx
1.0$\cite{science}. A related ratio has been formed for central to
peripheral collisions which produce low multiplicities, such as those in
$pp$ interactions.  Similar values were found\cite{phenix,star}.  The results
at $\sqrt{S_{NN}} = 130$ GeV and 200 GeV are in agreement with each other.

One possible explanation for the strong suppression is parton energy
loss in the medium produced in the collision \cite{vitev,jeon}.
However, other, more conventional nuclear effects must also be
considered.  Gold nuclei have a different isospin from the proton
reference.  Isospin is much less relevant for comparison of central
and peripheral ion collisions.  More importantly, the parton
distributions in nuclei are known to be different from those in bare
nucleons\cite{Arn}.  In their analysis, the experimenters noted that
this difference, referred to as nuclear shadowing, might affect
$R_{AA}$ but estimated that the change would be small.  Here, we give
quantitative estimates of the effects of nuclear shadowing and isospin
on $R_{AA}$ for charged pions, kaons and protons separately.  The
Cronin effect, which broadens the $p_T$ distributions in $pA$ relative
to $pp$ collisions, tends to increase $R_{AA}$ \cite{vitev}.  To
better illustrate the effects of shadowing and isospin alone, we do
not include the Cronin effect in our calculations.

We make a leading order (LO) calculation of minijet production to 
calculate the yield of high-$p_T$ partons\cite{usprc}.  The $p_T$ distribution
is \cite{Field}
\begin{eqnarray}
\frac{d\sigma_{AB}^h}{dp_T} = 2p_T \int_0^\pi \frac{d\theta_{\rm cm}}{\sin
\theta_{\rm cm}} \int dx_i \int dx_j f_{i/A}(x_i,Q^2) f_{j/B}(x_j,Q^2)
\frac{D_{h/k}(z_c)}{z_c} \frac{d\hat \sigma}{d \hat t} \, \,  
\end{eqnarray}
where $x$ is the fraction of the hadron momentum carried by the interacting
parton and $Q$ is the momentum scale of the interaction.
The integral over center-of-mass scattering angle $0 \leq \theta_{\rm cm} \leq 
\pi$ corresponds to an integral over all rapidities.
The $2 \rightarrow 2$ minijet cross sections 
are given by $d\hat \sigma/d \hat t$.
The fragmentation functions, $D_{h/k}(z_c)$, are the probability for the
production of hadron $h$ from parton $k$ \cite{fragment}.  
The parton densities, $f_{i/A}(x_i,Q^2)$, in a nucleus are $f_{i/A}(x_i) =
R_A^i(x_i)(Z_A f_{i/p}(x_i) + N_A f_{i/n}(x_i))/A$ where $Z_A$ and $N_A$ are
the proton and neutron numbers in nucleus $A$.
We use the MRST LO proton parton distribution functions for isolated
nucleons\cite{MRST} and the EKS98 parameterization of the shadowing function
$R_A^i$ \cite{Eskola}.  We take $Q^2 = p_T^2$.

In our calculations, we neglect higher-order corrections, using LO
parton densities, fragmentation functions, and shadowing
parameterizations.  Since $R_{AA}$ is a ratio, the higher-order
corrections should largely cancel out.  The produced quarks,
antiquarks and gluons are fragmented into charged pions, kaons and
protons using LO fragmentation functions fit to $e^+ e^-$ data
\cite{fragment}.  The final-state hadrons are assumed to be produced
pairwise so that $\pi \equiv (\pi^+ + \pi^-)/2$, $K \equiv (K^+ +
K^-)/2$, and $p \equiv (p + \overline p)/2$.  Any baryon asymmetry in
the initial state then has no effect on the baryon composition of the
final state.  The produced hadrons follow the parent parton
directions.  The $Q^2$ evolution is modeled using $e^+e^-$ data at
several different energies.  These fragmentation functions were also
compared to $p \overline p$, $\gamma p$ and $\gamma\gamma$ data.
After some slight scale modifications\cite{fragment2} they were able
to fit all the $h^-$ data.  However, there are significant
uncertainties in fragmentation when the leading hadron takes most of
the parton momentum \cite{fragment3}.

Shadowing is assumed to depend only on the parton momentum fraction
$x$, the momentum scale $Q^2$, the parton flavor and the nuclear mass
number $A$.  The EKS98 parameterization evolves each parton type
separately for $2.25 \leq Q^2 \leq 10^4$ GeV$^2$. In the momentum
range relevant to RHIC, $R_A^i$ for quarks and antiquarks are based on
nuclear deep-inelastic scattering data.  There is very little direct
data on the nuclear gluon distribution so that gluon shadowing is
primarily based on the $Q^2$ evolution of the nuclear structure
functions.  The gluon density shows significant antishadowing for
$0.1<x<0.3$ while the antiquark densities are shadowed in this region.
For $0.3<x<0.7$, there is significant suppression for all partons, the
EMC effect, while for $x<0.07$, there is also significant suppression.

We do not include inhomogeneous (spatially varying) modifications
\cite{inhomog}.  Shadowing should be largest near the core of the
nucleus and reduced near the surface.  In very peripheral collisions,
shadowing effects should be sufficiently reduced for $pp$ collisions
to be a reasonable model of the most peripheral events.  For the most
central collisions, the deviation from $R_{AA}=1$ differs by less than
1\% from than that calculated with homogeneous shadowing.

\begin{figure}
\setlength{\epsfxsize=0.7\textwidth}
\setlength{\epsfysize=0.5\textheight}
\centerline{\epsffile{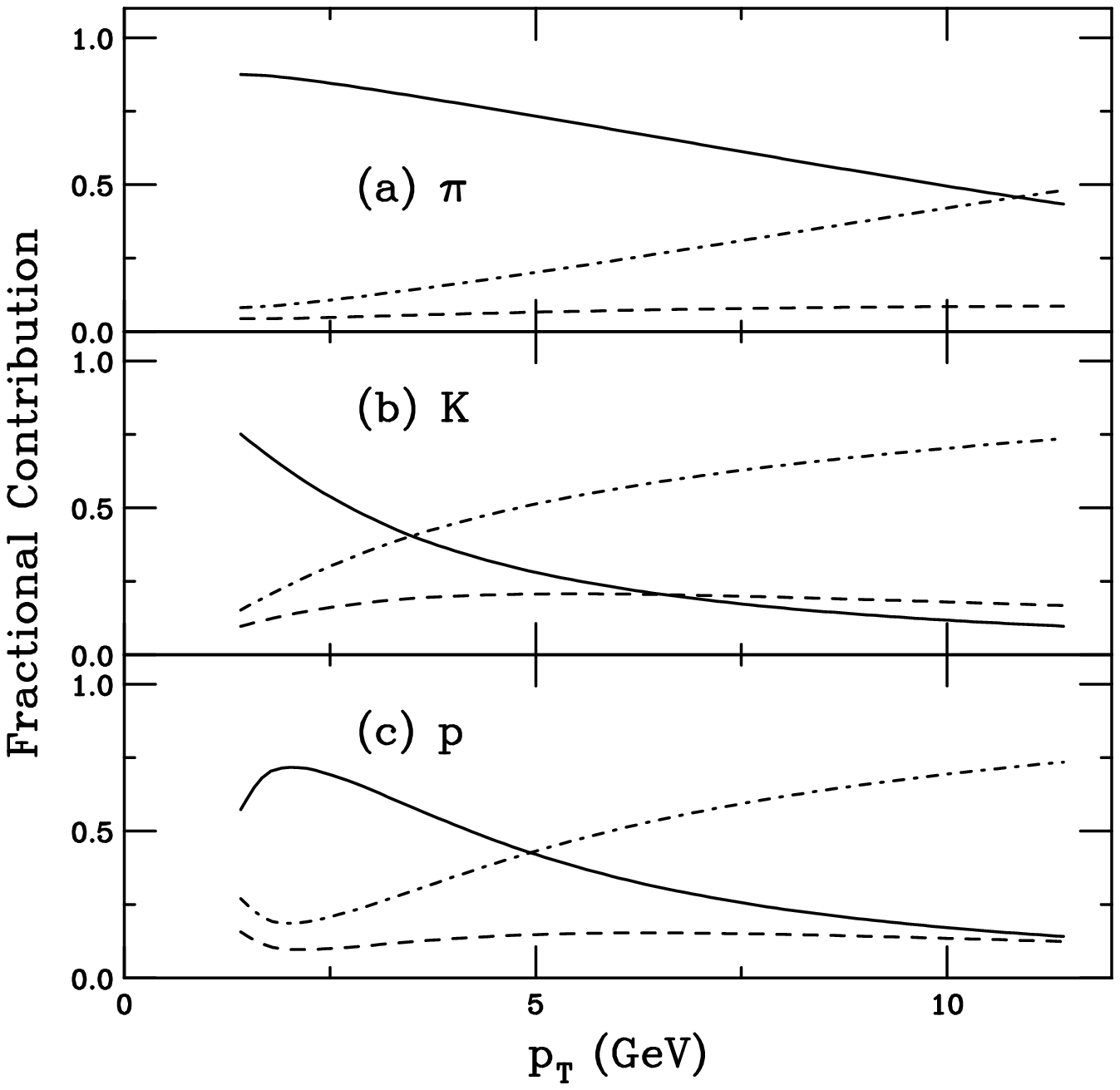}}
\caption{Fractional contribution of gluons (solid curves), quarks (dot-dashed
curves) and antiquarks (dashed curves) for (a) pion, (b) kaon, and (c) proton
production in $pp$ collisions at $\sqrt{S} = 200$ GeV as a function of $p_T$.}
\label{fig:frac}
\end{figure}

Minijets come from quarks, antiquarks and gluons produced in the hard
scattering.  The fragmentation function determines how these partons
become final-state hadrons.  Figure~\ref{fig:frac} shows the
percentage of charged pions, kaons and protons produced by quarks, antiquarks
and gluons as a function of $p_T$ at central rapidities, $|y| \leq 1$,
for $pp$ interactions at $\sqrt{S} = 200$ GeV.  Pion production
is dominated by gluons up to $p_T \sim 10$ GeV/$c$.  At higher
$p_T$, most of the pions come from quarks.  For kaons and protons,
the crossover between gluon and quark dominance occurs at lower $p_T$,
$p_T\sim 3.5$ GeV/$c$ for kaons and $\sim 5$ GeV/$c$ for protons.  At
large $p_T$, $\sim 75$\% of kaons and protons are produced by quarks.
At larger rapidities and low $p_T$,
the relative quark and antiquark contributions increase
for kaons and protons.  In fact, the gluon contribution
to kaon production becomes smaller than the quark and antiquark
contributions.

\begin{figure}
\setlength{\epsfxsize=0.7\textwidth}
\setlength{\epsfysize=0.4\textheight}
\centerline{\epsffile{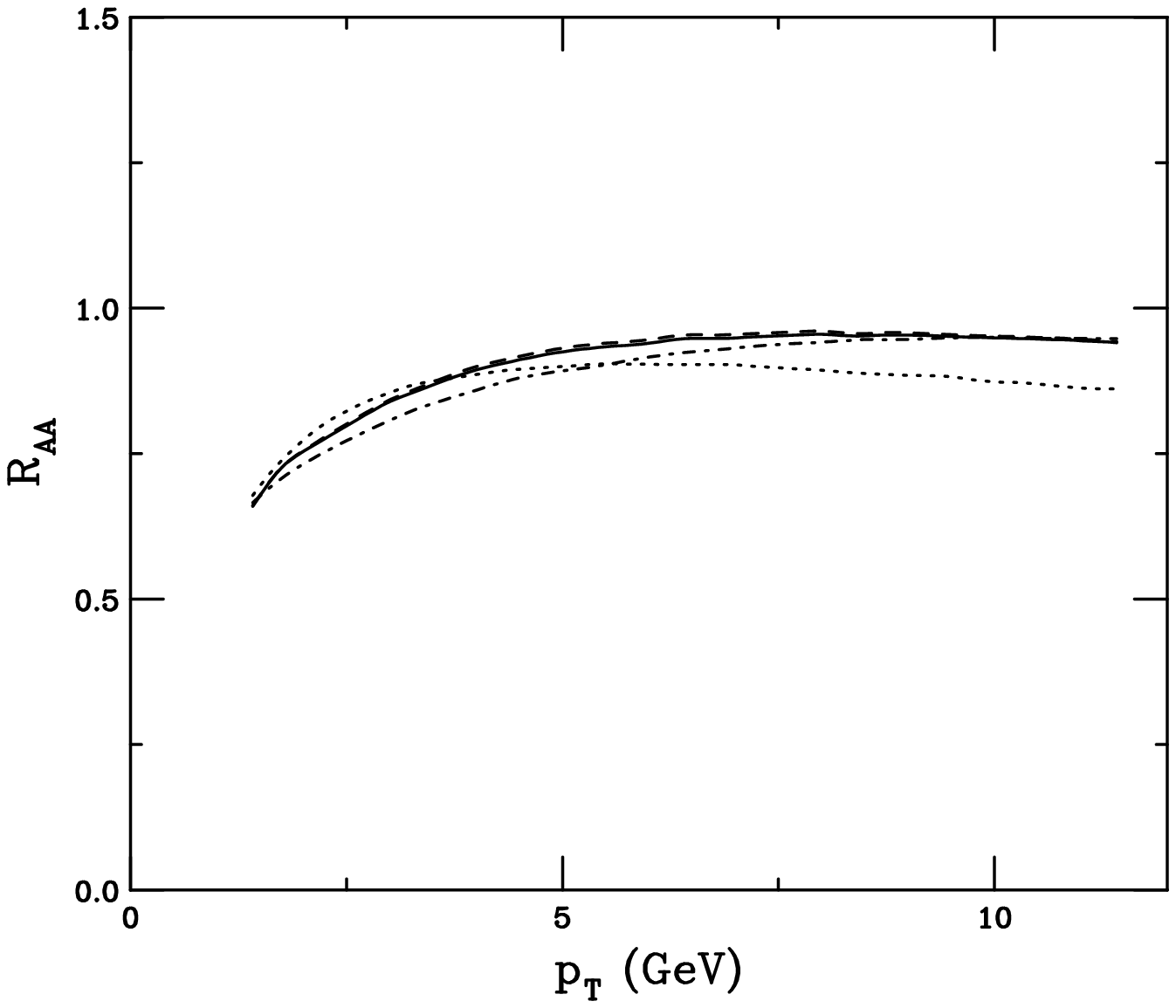}}
\caption{$R_{AA}$ for pions (dashed curve), kaons (dot-dashed curve),
protons (dotted curve) and the average over all hadrons (solid
line) for gold-gold collisions at $\sqrt{S_{NN}} = 200$ GeV 
as a function of $p_T$.  Spatially averaged shadowing is
used.}
\label{fig:raa}
\end{figure}

\begin{figure}
\setlength{\epsfxsize=0.7\textwidth}
\setlength{\epsfysize=0.4\textheight}
\centerline{\epsffile{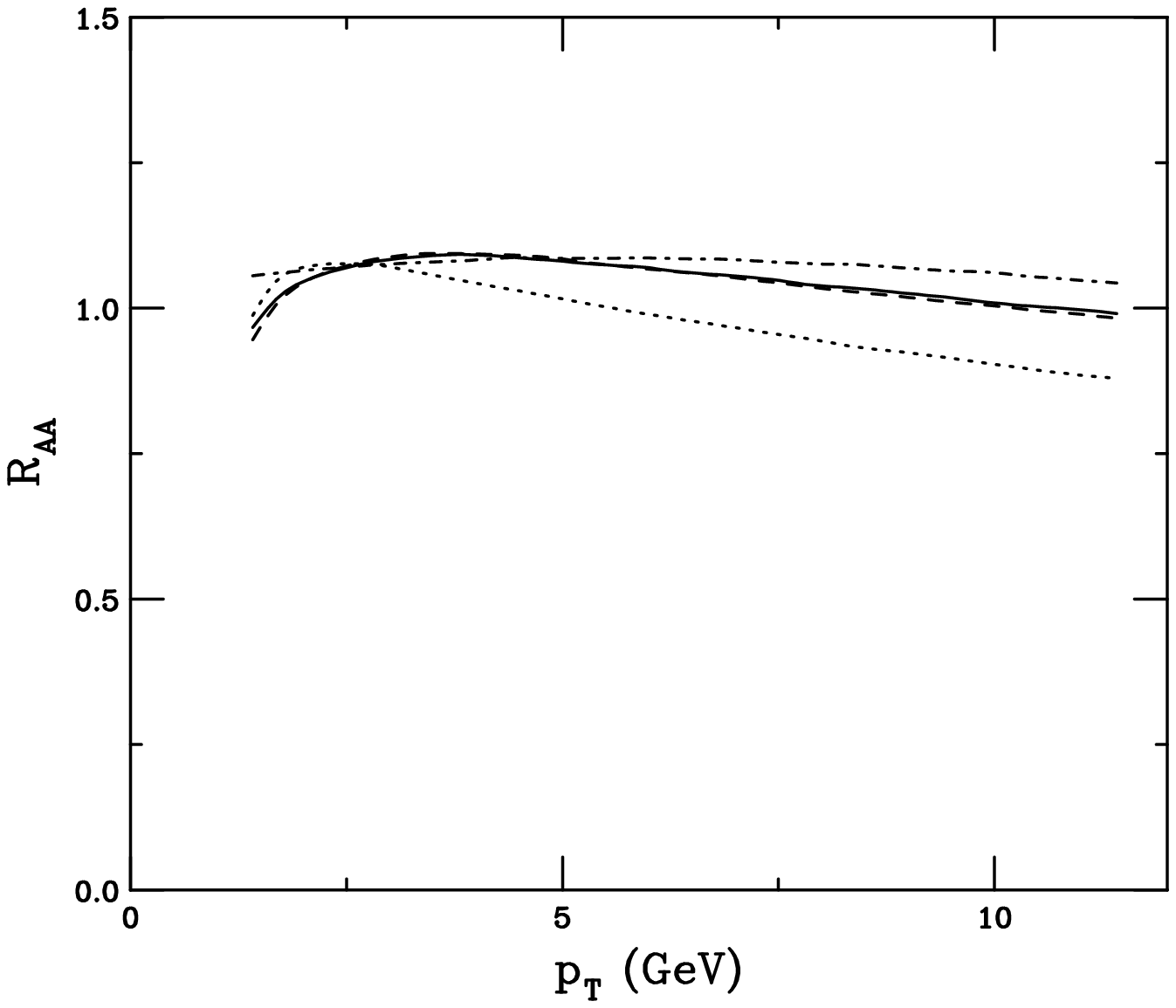}}
\caption{$R_{AA}$ for pions (dashed curve), kaons (dot-dashed curve),
protons (dotted curve) and the average over all hadrons (solid
line) for gold-gold collisions with $|y| \leq 1$ at $\sqrt{S_{NN}} = 200$ GeV 
as a function of $p_T$.  Spatially averaged shadowing is
used.}
\label{fig:raay}
\end{figure}

Figures~\ref{fig:raa} and \ref{fig:raay} compare the calculated
$R_{AA}$ integrated over all rapidities to that restricted to $|y|\leq
1$.  The region $|y| \leq 1$ roughly matches the experimental
acceptances.  The results shown in Figs.~\ref{fig:raa} and
\ref{fig:raay} probe different $x$ regions, as described below.
Without fragmentation, $x_{i,j} = (p_T/\sqrt{S_{NN}})(e^{\pm y_1} +
e^{\pm y_2})$ where $y_1$ and $y_2$ are the parton rapidities.  When
$y_1 = y_2 =0$, $x_i = x_j = 2p_T/\sqrt{S_{NN}}$.  For illustration,
we assume that the leading hadron of the minijet carries half the
parent parton momentum, increasing $x_i$ by a factor of two.  Then the
region $2 \leq p_T \leq 10$ GeV corresponds to $0.04 \leq x_i \leq
0.2$.  In this $x$ range, the EKS98 parameterization has valence quark
and gluon antishadowing and sea quark shadowing.  The shadowing
modifications decrease with $Q^2$.  Thus the strongest effects are at
low $p_T$ but the overall effect is small.  Away from $y=0$, $x_i$ and
$x_j$ are different.  One $x$ will decrease into the low $x$ shadowing
region while the other will increase into the EMC region.  There are
stronger modifications in both these regions so that shadowing has a
bigger effect in the broader rapidity region.  This is most clearly
seen for low $p_T$ in Fig.~\ref{fig:raa} where shadowing reduces
$R_{AA}$ $\approx 30$\% for all species.  At higher $p_T$ the effects
are smaller and only the proton curve comes close to the data.  The
large suppression observed at RHIC for mesons is not seen here.

Figure~\ref{fig:raay} shows that for $|y| \leq 1$, shadowing is only
a few percent effect due to the restricted $x$ regions.  The
composition changes, with charged kaons slightly enhanced and protons slightly
suppressed compared to charged pions. This difference is due to isospin.  The
fragmentation functions assume $u$ and $s$ quarks and antiquarks
fragment identically to charged kaons \cite{fragment} so that at large $p_T$,
charged kaon production is favored in $AA$ collisions relative to $pp$.  For
neutral kaons, the situation is reversed.  On
the other hand, a $u$ is twice as likely to produce a proton than a
$d$ \cite{fragment} so that proton production is favored in $pp$
interactions.  The isospin effect is larger than the nuclear
modifications on kaons and protons for $p_T> 7.5$ GeV/$c$.  The
dominance of pion production by gluons and the assumption that $u$ and
$d$ are equally likely to produce charged pions \cite{fragment} leads to a
negligible isospin effect for all $p_T$.  Thus $R_{AA}$ for pions is
influenced only by the nuclear modifications.   A similar conclusion was
reached elsewhere \cite{jeon}.

Note that in both Figs.~\ref{fig:raa} and \ref{fig:raay}, the total $R_{AA}$
closely follows the pion result.  This is due to the relative fragmentation
yields: the pion yield is significantly higher than the proton yield at all
$p_T$, in contradiction with the data \cite{phenix}.  
Reference \cite{fragment3} also showed that the $p/\pi$ ratio was
underpredicted by the fragmentation functions \cite{fragment}.  

Other models may predict larger shadowing effects on $R_A^i$.  For
example, the shadowing model of Ref.~\cite{strikman}, based on
diffractive data from HERA, predicts large gluon shadowing for
$x<0.01$, $R_A^g \approx 0.3$ at $x=0.001$. They also predict
significant antishadowing for $x>0.03$, comparable to EKS98.  The
gluon shadowing grows rapidly when $x < 0.01$.  This model should lead
to significant reduction in $R_{AA}$ for large rapidities at RHIC.
However, in the region $|y|\leq 1$, $x\approx 0.02-0.1$, nuclear
effects are rather small and better constrained by data.  Thus
$R_{AA}$ should not be significantly model dependent.

Models with strong gluon saturation and/or classical gluon fields,
like the colored glass condensate\cite{colorglass}, predict very
different nuclear gluon densities.  However, $p_T > Q_s$, the
saturation scale predicted at RHIC \cite{colorglass}.  In addition, at
midrapidity, the relevant $x$ values are not so small and are in a
region where fixed-target data constrains the quark and antiquark
modifications.  Theoretical analysis indicates that the gluon
distributions are also rather well constrained in the RHIC kinematic
region\cite{jamal}. These studies strongly limit the possible effect
on $R_{AA}$ near midrapidity.

At the LHC, where $\sqrt{S_{NN}} = 5.5$ TeV, the situation for lead on
lead collisions will be very different.  A hadron with $p_T = 5$
GeV/$c$ (or a parton with $p_T = 10$ GeV/$c$) corresponds to $x
\approx 0.002$, and shadowing is significant, reducing $R_{AA}$ below
1.  Away from $y=0$, even smaller $x$ values are probed.  Gluon
dominance of minijet production extends to larger $p_T$, considerably
reducing the effect of isospin.  Thus nuclear modifications will be
significant at the LHC.

In conclusion, nuclear shadowing cannot explain a significant fraction
of the observed suppression of high-$p_T$ particles at RHIC.  With the
EKS98 parameterization and the nuclear isospin, at midrapidities $1.0
< R_{AA} < 1.1$ for charged mesons and $R_{AA} \approx 1$ for protons.
Without the restriction $|y| \leq 1$, shadowing and isospin have a
bigger effect on $R_{AA}$, with $R_{AA}\approx 0.7$ at $p_T=2$ GeV/c,
rising to 0.8 at $p_T=10$ GeV/c.  Models with stronger nuclear effects
may further reduce $R_{AA}$ when all rapidities are considered.
However, at mid-rapidity, no large effect on $R_{AA}$ is expected from
any model that reproduces the existing lepton scattering data.

We thank K.J. Eskola for providing the shadowing routines.  This work
was supported in part by the Division of Nuclear Physics of the Office
of High Energy and Nuclear Physics of the U. S. Department of Energy
under Contract Number DE-AC03-76SF0098.

\newpage
\newpage

\end{document}